\documentclass[twocolumn,showpacs,amsmath,amssymb]{revtex4}

\usepackage{graphicx}
\usepackage{dcolumn}
\usepackage{bm}

\begin{document}

\title{\center{Wave functions in the neighborhood of a toroidal surface;\\
hard vs. soft constraint}}

\author{Mario Encinosa }
\author{Lonnie Mott}%

\author{Babak Etemadi}
\affiliation{ Florida A\&M University,Department of Physics,
Tallahassee FL 32307}


\begin{abstract}
The curvature potential arising from confining a particle
initially in three-dimensional space onto a curved surface is
normally derived in the hard constraint $q \rightarrow 0$ limit,
with $q$ the degree of freedom normal to the surface.  In this
work the hard constraint is relaxed, and eigenvalues and wave
functions are numerically determined for a particle confined to a
thin layer in the neighborhood of a toroidal surface. The hard
constraint and finite layer (or soft constraint) quantities are
comparable, but both differ markedly from those of the
corresponding  two dimensional system, indicating that the
curvature potential continues to influence the dynamics when the
particle is confined to a finite layer. This effect is potentially
of consequence to the modelling of curved nanostructures.
\end{abstract}

\maketitle

\section{\label{sec:level1}Introduction} The existence of a potential
$V_C$ in the Schrodinger equation which stems from constraining a
particle to a one or two-dimensional surface embedded in three
dimensions has a long history
\cite{jenskoppe,dacosta1,dacosta2,exnerseba,matsutani,burgsjens}.
The manifestations of $V_C$ have been investigated through formal
and numerical means
\cite{duclosexner,bindscatt,goldjaffe,ouyang,popov,midgwang,clarbrac,schujaff,ee1,
ee2,ieee}, motivated recently in part by the sophistication with
which nanostructures can be fabricated. The physics of objects
with novel geometries is increasingly relevant to the modelling of
real devices, hence substantial effort has been directed towards
understanding the physics of bent tubes and wires, as well as more
complicated shapes
\cite{lin,chapblic,qu,nils,mott,halberg1,halberg2}.

Consider a surface $\Sigma(u,v)$ with $(u,v)$ surface coordinates
and define $q$ as the coordinate labelling the degree of freedom
normal to $\Sigma(u,v)$. Generally $V_C$ (as detailed in the
following section) is derived by imposing a hard constraint on the
particle wherein a $q \rightarrow 0$ limit is taken along with a
wave function re-scaling such that the norm is preserved. Here,
 rather than imposing a hard constraint, the
 particle will be confined to a thin
layer in the neighborhood of a toroidal surface. The extent to
which the hard constraint mirrors the more physically realizable
soft constraint is then determined by calculating some low-lying
eigenvalues and eigenfunctions of the system.

There are several reasons to investigate these ideas with a
torodial structure: \vskip 6pt \noindent 1. The symmetry of the
torus reduces computational intensiveness, but, because a torus
has non-trivial mean and Gaussian curvatures,  curvature effects
remain important \cite{encmott}. \vskip 6pt \noindent
 2. The spectrum and eigenfunctions for a particle on a toroidal
surface have been determined \cite{fpl} , so comparisons can be
made between the finite layer system and the two dimensional
system both with and without $V_C$ present.

\vskip 6pt \noindent 3. Toroidal structures have been fabricated
and calculations addressing their transport
\cite{shea,sano,latge,sasaki} and magnetic \cite{liu} properties
performed. Toroidal structures are novel because unlike a bulk
sample, conductivity through the device is anticipated to be
dominated by azimuthal modes. \vskip 6pt

The remainder of this paper is organized as follows: in section
II, $H_q$, the Hamiltonian for a particle near a toroidal surface
is derived. The hard constraint $q  \rightarrow 0$ limit of that
Hamiltonian is then taken; under the requirement that the norm of
the wave function be preserved, $H_C$ obtains.  Finally, the ab
initio $q = 0$ Hamiltonian $H_0$ is written. In section III the
computational method employed to generate eigenvalues and wave
functions is presented. Section IV gives results and section V is
reserved for conclusions.

\section{The   toroidal  Schrodinger \ equations}
To restate, there are three Hamiltonians relevant to this work:
\vskip 6pt \noindent 1. $H_q$ will be the Hamiltonian for a
particle allowed to move in a thin layer normal to $T^2$, where
again the normal degree of freedom will be labelled by $q$.
\vskip6pt \noindent 2. $H_C$ will be the Hamiltonian derived from
$H_q$ after imposing the $q \rightarrow 0$ hard constraint, and
\vskip 6pt \noindent 3. $H_0$ the Hamiltonian for a particle
restricted ab initio to $T^2$, i.e., $q = 0$ at the onset of the
derivation of $H_0$. \vskip 6pt It is best to begin with the most
general case, that of $H_q$, and later take the appropriate limits
to obtain $H_C$ and $H_0$.

Points near a toroidal surface of major radius $R$ and minor
radius
 $a$ may be parameterized in terms of  cylindrical coordinate unit
 vectors and a vector $\mathbf{\hat{n}}$ normal to the surface
 by \cite{fpl}
$$
 {\bf r}(\theta,\phi,q)=(R + a \ {\rm cos}  \theta )\mathbf{\hat{\rho}} +a\  {\rm sin}
\theta\mathbf{\hat {k}} + q\mathbf{\hat{n}}. \eqno(1)
$$
 Applying $d$ to Eq.(2)
gives
$$
d{\bf r}= (a+q)d\theta\mathbf{ \hat{\theta}}+(R + (a+q) \ {\rm cos}
\theta)d\phi\mathbf{ \hat{\phi}}+dq\mathbf{ \hat{n}} \eqno(2)
$$
with $\mathbf{\hat{n}} \equiv \mathbf{ \hat{\phi}} \ {\rm x}\  \mathbf{
\hat{\theta}}$, and $\mathbf{\hat{\theta}} =-\rm sin \theta
\mathbf{\hat{\rho}}+\rm cos \theta \mathbf{\hat {k}}$.
 The metric elements $g_{ij}$ can be read off of
$$
d{\bf r}\cdot d{\bf r}=(a+q)^2 d\theta^2+
                (R+(a+q){\rm cos}\theta)^2d\phi^2 + dq^2
\eqno(3)
$$
and the Laplacian  derived from
$$
\nabla^2= g^{-{1 \over 2}}{\partial \over \partial q^i} \bigg [
g^{1 \over 2}\ g^{ij}{\partial \over \partial q^j} \bigg ].
\eqno(4)
$$
Setting $a_q = a+q$ and $F_q = R+ (a+q) \rm cos \theta$ yields the
 $H_q$ Schrodinger equation
 \begin{widetext}
$$
 {1 \over {a_q^2}}{\partial^2 \psi \over \partial \theta^2} -
  {{\rm sin} \theta \over a_q F_q} {\partial \psi \over \partial \theta}
+{1 \over F_q^2}{\partial^2 \psi \over
\partial \phi^2} +  2h {\partial \psi \over \partial q}
+ {\partial^2 \psi \over \partial q^2} - 2V_n(q) + 2E\psi =0
\eqno(5)
$$
\end{widetext}
with $h$ the mean curvature  given by
$$
h \equiv {1 \over 2}(k_1+k_2) = {1 \over 2}\bigg[{1 \over a_q}+
{{\rm cos}\theta \over F_q}\bigg]. \eqno(6)
$$
It is convenient to also define the  Gaussian curvature $k$,
$$
k \equiv k_1k_2 = {1 \over a_q} {{\rm cos}\theta \over F_q}.
\eqno(7)
$$

 To derive the Schrodinger equation appropriate to
$H_C$, $V_n(q)$ must be chosen
 to drive the particle arbitrarily close to the
$q = 0$ limit{\cite{jenskoppe,dacosta1,dacosta2}}. As $q
\rightarrow 0$ the wave function is expected to decouple into
surface and normal parts as
$$
\psi(\theta,\phi,q) \rightarrow \chi_s(\theta,\phi)\chi_n(q).
\eqno(8)
$$
Conservation of the norm must be preserved leading to
\cite{jenskoppe,dacosta1,dacosta2,matsutani2,kaplan}
$$
|\psi|^2 WdSdq=|\chi_s|^2|\chi_n|^2dSdq \eqno(9)
$$
or
$$
\psi =\chi_s \chi_n W^{-{1\over2}} \eqno(10)
$$
 where $W=1+2qh+q^2k$ and $dS$ the surface measure.
Performing the differentiations and letting $q\rightarrow 0$ gives
the pair of equations
$$
 {\partial^2 \psi \over \partial \theta^2} -
 {\alpha \ { \rm sin}\ \theta \over F}
 {\partial \psi \over \partial
 \theta}
+{{\alpha}^2 \over F^2}{\partial^2 \psi \over
\partial \phi^2} -2a^2 V_C +\beta \psi =0, \eqno(11)
$$

$$
-{1 \over 2}{\partial^2  \chi_n \over \partial q^2} + V_n(q)
\chi_n =E_n \chi_n \eqno(12)
$$
 with $\alpha = a/R, \beta = 2Ea^2$ and $F=1+\alpha \ \rm cos\theta$.
 The curvature potential $V_C$ is
$$
V_C = -{1\over 8 a^2} {1\over  F^2}. \eqno(13)
$$

Making the standard ansatz for the azimuthal part of the
eigenfunction $\chi(\phi) = exp \ [im\phi]$  reduces Eq. (11) to
$$
  {\partial^2 \psi \over \partial \theta^2} -
 {\alpha \ {\rm sin}\ \theta \over [1 + \alpha \   \cos\theta]}{\partial \psi \over \partial \theta}
-{(m^2 \alpha^2- {1\over4}) \over [1 + \alpha \ \cos\theta]^2}\psi
+\beta\psi = 0. \eqno(14)
$$
Eq. (14) is the Schrodinger equation that corresponds to $H_C$. It
is the analog to Eq. (5) wherein the $q$ dependence has decoupled
from the the angular part of the kinetic energy operator and a
curvature potential $V_C$ results from insisting upon conservation
of the norm.

The Hamiltonian $H_0$ for a particle that lives on the surface may
be obtained by the method employed to derive $H_q$ by setting $q =
0$ in  Eq. (1) from which
$$
d{\bf r}\cdot d{\bf r}=a^2 d\theta^2+
                (R+a\ {\rm cos}\theta)^2d\phi^2.
\eqno(15)
$$
The resulting expression is simple; $H_0$ is Eq. (11)  with $V_C$
omitted \cite{fpl}. It should be emphasized that for more
complicated surfaces the kinetic energy operator will have terms
depending on the surface curvature not present here because of the
azimuthal symmetry of the torus \cite{ee1}.
 The normalization of an
eigenfunction is determined by
$$
\int^{q_f}_{q_i} \int^{2\pi}_0\int^{2\pi}_0 \psi^*(q,
\theta,\phi)\psi(q,\theta,\phi) M(\theta, q) d\theta d\phi dq= 1.
\eqno(16)
$$
with
$$
M(\theta,q) = a_q F_q \eqno(17)
$$
when $q \neq 0$. Wave functions obtained from $H_{0,C}$ are
normalized with $M(\theta,0)$ and the $q$ integration omitted.

\section{Computational method}

The goal is to obtain eigenvalues/functions of $H_q$ that can be
compared to those of $H_C$ and $H_0$. A procedure for determining
the low-lying eigenvalues and eigenfunctions of $H_0$ has been
given in \cite{fpl} and applied to $H_C$ in \cite{encmott} so the
focus here may be placed on the method employed for solving Eq.
(5). In the $q \rightarrow 0$ limit the surface solutions are
independent of the specific choice of $V_n(q)$, but for finite $q$
a form or forms for $V_n(q)$ must be settled upon. Two convenient
choices are hard wall confinement with the walls  at $\pm {L / 2}$
and an oscillator potential $V_n(q)= \omega^2 q^2/2$.

The main complication in solving Eq. (5) ensues from the
integration measure for the geometry described by Eq. (3), which
precludes adopting a simple basis set of trignometric functions in
$\theta$ since they are not in general orthogonal over
$M(\theta,q)$. It should also be noted that for finite $q$ it is
not possible to recover orthogonality by re-scaling the basis
states because the resulting Hamiltonian matrix is not Hermitian.
These difficulties can be avoided if one performs a two-variable
Gram-Schmidt procedure.

Because the interest here is focused on issues other than
generating a large number of eigenfunctions, two basis functions
in $q$ and three in $\theta$ were used. This is a reasonable
ansatz;
 the $q$ motion will produce an energy
spectrum with much larger spacing than the $\theta,\phi$ motion,
 so two functions in $q$ are  sufficient to insure nothing is
missed. Additionally, it has been shown previously \cite{fpl} that
only a few trigonometric functions in general are necessary to
accurately describe an eigenfunction on $T^2$.

 For hard walls  the
basis functions with $n = (0,1,2)$ are
$$
\Phi^{0n}_{hw}={\rm cos}({\pi q \over L}){\rm cos}(n\theta)
\eqno(18)
$$
$$
\Phi^{1n}_{hw}={\rm sin}({2\pi q \over L}){\rm cos}(n\theta)
\eqno(19)
$$
 and for  oscillator confinement
$$
\Phi^{0n}_{osc}=e^{-\omega q^2}{\rm cos}(n\theta) \eqno(20)
$$
$$
\Phi^{1n}_{osc}=e^{-\omega q^2}H_1(\sqrt \omega q){\rm
cos}(n\theta).
 \eqno(21)
$$
For each case the six states computed from the Gram-Schmidt
procedure  are employed to construct the matrix
$$
H^{ijmn}_{hw,osc}= \big<\Phi^{im}_{hw,osc}|H_q|\Phi^{jn}_{hw,osc}
\big > \eqno(22)
$$
that yields  eigenvalues and wave functions.

\section{Results}
Toroidal radii $R = 500 \AA $ and $a = 250 \AA$ were chosen on the
order of structures that have been synthesized \cite{garsia,
lorke,zhang}, and surface layer widths as set by $L, \omega$
 within realistic values for confinement regions.
It should be emphasized that the results which follow are very
representative; the trends exhibited below were found to obtain
for larger values of $R, a,$ and $L$ as well as for $m \neq 0$ and
negative parity states.

In table I the spectra for hard wall and oscillator confinement
potentials with $L = 25, 10$ $ \AA$ and $\omega = .05,.1 $
$\AA^{-2}$ respectively are shown.
 The dimensionless eigenvalues
$\beta_i$ are found from subtracting the $q$ degree of freedom
energy ($\pi^2/2L^2$ or $\omega /2$) from the eigenvalues found
from the Hamiltonian matrix defined through Eq. (22) and
multiplying by $2a^2$. The $\beta_i$ are compared to those found
in \cite{encmott} where $V_C$ was included in the $T^2$
Hamiltonian and in \cite{fpl} where it was not. These results
indicate the soft constraint quantities are relatively insensitive
to differing $L$ and $\omega$, and are better matched by the
spectra of \cite{encmott}. In tables II and III the ground and
first excited state wave functions
 for the six cases described above are shown.
 The results illustrate that hard constraint eigenvalues and
eigenfunctions  are very good approximations to the physically
realistic soft constraint values, at least for cases where the
length scale that determines the surface energies of the system is
near the curvature length scale of the device. Here that scale is
set by the minor radius $a$; however, in general as the length
scale that sets local curvature
 becomes small, $V_C$ increases such that $\big < i|V_C|j \big >$ matrix elements may become
  comparable to the largest
energy in the system. For a disc or strip structure  the scales
can be very different. In the case of a disk for example, the
energy scale is set by the radius of the disk, but a bump or
ripples can be placed on the disc at much smaller scales
\cite{ee2,ieee}. Although the results here are relatively
independent of whether hard wall or oscillator confinement was
used, it was found that some care must be taken with the choice of
$V_n(q)$. If instead of using hard walls at $\pm L/2$, the walls
are placed at $0$ and $L$,  agrement with the hard constraint
spectrum is lessened, though by only of order ten percent in both
the eigenvalues and wave function expansion coefficients. A
possible explanation for this is the ${\rm sin} (n\pi q/L)$
functions always vanish on the $q = 0$ surface so that some terms
that multiply curvature functions are zero there.

The basis set expansion employed here  comprises two functions in
the $q$ degree of freedom; for the sake of brevity, only angular
eigenvalues/eigenfunctions which belong to the $q$ ground state
wave function have been reported. The surface states that
correspond to excited normal modes lie much higher in energy than
the low-lying surface excitations dealt with here, but may prove
important to device modelling as the $q$-motion becomes more
diffusive.

\section{Conclusions}

 The main result of this paper is the good agreement between the
 low-lying
spectra and eigenfunctions resulting from $H_0$ and those that
emerge from $H_q$, and, as importantly, the relative disagreement
that the eigenfunctions of $H_q$ display when contrasted to those
of $H_0$. If a two-dimensional approximation is to be adopted for
 curved nanostructures, the results here indicate
that for cases where the normal excitations are unimportant the
 physics would be better captured with $H_0+V_C$ than with
$H_0$.

\begin{table}
\caption{Ground, first and second excited state eigenvalues
$\beta_i$ for the six Hamiltonians relevant to this paper with $R
= 500 \AA$ and $a = 250\AA$. }
\begin{center}
\begin{tabular}{|c|c|c|c|c|c|c|}
\hline
\ \   & $L =25 \AA$ & $L = 10 \AA$ & $\omega = .05\AA^{-2}  $& $\omega = .1 \AA^{-2}$&
Ref.\protect\cite{encmott} & Ref.\protect\cite{fpl} \\
\hline
$\beta_0$ & -.3405 & -.3406 & -.3489 & -.3488 & -.3511 & .0 \\
$\beta_1$ & .6618  & .6610 & .6515 & .6446 & .6386 & 1.1223\\
$\beta_2$ & 3.7919 & 3.7886 & 3.7800 & 3.7876 & 3.6529 & 4.0520\\
 \hline
\end{tabular}
\end{center}
\end{table}

\begin{table}
\caption{Ground state wave functions; coefficients are normalized
to the constant term in the series to facilitate comparisons.
Terms not shown are at least an order of magnitude smaller than
those given.}
\begin{center}
\begin{tabular}{|l|l|}
\hline
  & \qquad \qquad \ \ $\psi_0(\theta,q)$\\
\hline

$L=25 \AA$ & $(1- .3676 \ \rm cos \theta+.0693 \  \rm cos2 \theta){\rm cos {\pi q \over 25}}$\\
$L=10\AA $ & $(1- .3675 \ \rm cos \theta+.0693 \ \rm cos2 \theta){\rm cos {\pi q \over 10}} $ \\
$\omega = .05\AA^{-2}$ & $(1- .3580 \ \rm cos \theta+.0669 \ \rm cos2 \theta )e^{-.025q^2}$ \\
$\omega = .1\AA^{-2}$ & $(1- .3567 \ \rm cos \theta+.0654\  \rm \ cos2 \theta)e^{-.05q^2} $  \\
Ref.\protect\cite{encmott}  & $1- .3679 \ \rm cos \theta+.0784 \ \rm cos2 \theta$  \\
Ref.\protect\cite{fpl} & $1$  \\
 \hline
\end{tabular}
\end{center}
\end{table}

\begin{table}
\caption{First excited state wave functions; coefficients are
normalized to the dominant $\rm cos\theta$ term in the series to
facilitate comparisons. Terms not shown are at least an order of
magnitude smaller than those given.}
\begin{center}
\begin{tabular}{|l|l|}
\hline
  & \qquad \qquad \ \ $\psi_1(\theta,q)$\\
\hline

$L=25\AA $ & $(-.0842+ \rm cos \theta-.1369 \ \rm cos2 \theta){\rm cos {\pi q \over 25}}$\\
$L=10\AA $ & $(-.0842+\rm cos \theta- .1370 \ \rm cos2 \theta){\rm cos {\pi q \over 10}} $ \\
$\omega = .05\AA^{-2}$ & $(-.0879 + \rm cos \theta-.1358 \ \rm cos2 \theta )e^{-.025q^2}$ \\
$\omega = .1\AA^{-2}$ & $(-.0877 +  \rm cos \theta-.1362 \ \rm cos2 \theta)e^{-.05q^2} $  \\
Ref.\protect\cite{encmott} & $-.0851 + \rm cos \theta - .1540 \ \rm cos2 \theta$  \\
Ref.\protect\cite{fpl} & $-.2500+ \rm cos \theta -.0820 \ \rm cos 2 \theta$  \\
 \hline
\end{tabular}
\end{center}
\end{table}

\newpage

\end{document}